\documentclass[aps,preprint]{revtex4}
\usepackage{epsfig}
\usepackage{graphics}
\usepackage{amssymb}
\usepackage{verbatim}
\usepackage{color}
\date{ }

\begin{document}
\title{Quantum transport through single and multilayer icosahedral fullerenes}
\author{Daniel~A.~Lovey} 
\author{Rodolfo~H.~Romero}
\email{rhromero@exa.unne.edu.ar}

\date{\today}

\begin{abstract}
We use a tight-binding Hamiltonian and Green functions methods to calculate the quantum transmission through single-wall fullerenes and bilayered and trilayered onions of icosahedral symmetry attached to metallic leads. The electronic structure of the onion-like fullerenes takes into account the curvature and finite size of the fullerenes layers as well as the strength of the intershell interactions depending on to the number of interacting atom pairs belonging to adjacent shells. 
Misalignment of the symmetry axes of the concentric iscosahedral shells produces breaking of the level degeneracies of the individual shells, giving rise some narrow quasi-continuum bands instead of the localized discrete peaks of the individual fullerenes. As a result, the transmission function for non symmetrical onions are rapidly varying functions of the Fermi energy. Furthermore, we found that most of the features of the transmission through the onions are due to the electronic structure of the outer shell with additional Fano-like antiresonances arising from coupling with or between the inner shells.
\end{abstract}
\maketitle
\section{Introduction}
Fullerenes, hollow clusters made up of carbon atoms bonded by sp$^2$ orbitals, have interesting conducting properties. The origin of such a behavior 
is their delocalized $\pi$ frontier molecular orbitals, 
what gives rise to a high conductance when an electric field is applied on the molecule, e.g. by an external potential bias. It has been discussed that free-electron and tight-binding (TB) models can capture the main features of the electronic transport through nearly spherical fullerenes \cite{Manousakis91, Mizorogi03}. The high conductivity of C$_{60}$ has lead to speculate on the possibility of considering it as a conducting spherical shell \cite{Amusia06}.
Particularly, several authors have studied the molecular junctions of the C$_{60}$ fullerene under different types of connections, such as, 
a substrate and a STM tip \cite{Paulsson99, Neel08}, one-dimensional leads \cite{Saffarzadeh08}, carbon nanotubes \cite{Shokri11}, gold clusters \cite{Bilan12} or break junctions \cite{Lortscher13}. 
Furthermore, the stability and strong hybridization of C$_{60}$ with metallic surfaces make it also a feasible anchoring group with high conductance \cite{Bilan12}. 
In the search of similar suitable molecular junctions, other larger icosahedral fullerenes C$_n$ from the same family $n=60 k^2$ ($k$ integer) have also been shown to be stable \cite{Yu09, Dunlap06}, while the conductance of others, 
such as C$_{20}$ or its complexes have also been explored \cite{Otani04, An09, Ji12}. On the other hand, by doping with boron and nitrogen, fullerene-based molecular junctions were found to have negative resistance \cite{Yaghobi11}. 

A number of methods have been applied to the study of energetics and stability of buckyonions, namely, icosahedral fullerenes encapsulated by larger ones \cite{Maiti93, Guerin97, Heggie97, Heggie98, Dodziuk00, Glukhova05, Baowan07, Enyashin07, Pudlak09, Xu08, Charkin13}.
Carbon nano-onions are interesting structures between fullerenes and multi-wall carbon nanotubes, having high thermal stability and chemical reactivity compared to CNTs. They also have the characteristic high contact area and affinity for noble metals, what make them interesting as anchoring groups for molecular electronics. 
For instance, it has also been shown that onion-like nanoparticles can be used as electrochemical capacitors, also called supercapacitors, with high discharge rates of up to three order of magnitud higher than conventional supercapacitors \cite{Pech10}.
The static polarizability, closely related to the response of the electronic charges to applied static fields, have been studied for onions formed by members of the icosahedral ${60 k^2}$ family, using both phenomenological models and first-principle methods \cite{Iglesias-Groth03, Zope08, Gueorguiev04}, showing their capability to partially screen static external electric fields.  The conductance of onion-like structures functionalized with sulfide-terminated chains has been measured between a gold substrate and a gold STM tip \cite{Sek13}.

The present work is aimed to study the electronic transmission of single-wall and multi-wall fullerenes weakly attached to metallic leads. 
In the next Section we discuss the TB 
model and the influence of the curvature and finite size of the layers on it, 
as well as the Green function-based method for the calculation of the transmission.
In Section III, we present our results for the dependence of the transmission function on the electron energy $T(E)$. 
We study how $T(E)$ is affected by the relative angular orientation of the shells, the number of intershell connections included in the TB 
model and the number of shells of the onions. Finally in Section IV, we summarize our conclusions on the systems studied.
\section{Model and calculation method}
\subsection{Single-wall fullerenes: curvature and finite size \label{one shell model}}
The Hamiltonian of a $n$-atom fullerene $C_n$ is described in the TB
approximation with one $\pi$ orbital per site
\begin{equation}
H_n = t_n \sum_{\langle ij\rangle} c_i^\dagger c_j + {\rm H.c.},
\end{equation}
where the summation runs on nearest neighbor atom pairs $\langle ij\rangle$, and the operator $c_i^\dagger$ ($c_j$) creates (annihilates) an electron in the $\pi$ orbital centered at the atom $i$ ($j$). The constant on-site energy has been taken as zero.

A single parameter $t_n$ is used for the hopping integral between nearest neighbor atoms for a given fullerene.
Although no bond dimerization effect is included, it has been shown that its sole effect is to slightly break some degeneracies in the spectrum with no major qualitative effect \cite{Manousakis91}. 
In order to take into account the effect of the curvature of the shell for the various fullerenes, we consider the
hopping parameter $t_n$ to be a function of the mean radius $R_n$ of the (nearly) spherical shell of $n$ C atoms and the mean inter-atomic distance $d_n$ \cite{Dresselhaus02}
\begin{equation}
t_n=t \left[1-\frac{1}{2}\left(\frac{d_n}{R_n}\right)^2\right].
\end{equation}
The value $t =-2.73$ eV is a suitable hopping for graphene and is chosen to correctly reproduce the HOMO-LUMO gap for C$_{60}$,  $E_g=-1.90$ eV, as obtained from DFT calculations \cite{Saito91}. Although the TB 
Hamiltonian only depends on the topology of the molecule (i.e., on the atoms bonded), the molecular geometry affects the hopping integral. 
\begin{table}
\begin{center}
\begin{tabular}{ccccc|ccccc}
\hline \hline 
C$_n$ & $d_n$ (\AA) & $R_n$ (\AA) & $t_n$ (eV) &&& C$_n$ & $d_n$ (\AA) & $R_n$ (\AA) & $t_n$ (eV) \\ \hline
C$_{60}$  & 1.43 &  3.54 & 2.51 &&& C$_{20}$  & 1.42 &  1.99 & 2.04 \\
C$_{240}$ & 1.42 &  7.05 & 2.68 &&& C$_{180}$ & 1.42 &  6.10 & 2.66 \\
C$_{540}$ & 1.42 & 10.58 & 2.71 &&& C$_{500}$ & 1.42 & 10.18 & 2.70 \\
\hline \hline
\end{tabular}
\caption{\label{d R t} Mean interatomic C-C distances, mean radii and hopping integral for the single-wall icosahedral fullerenes of the families $60k^2$ and $20k^2$ ($k$ integer).}
\end{center}
\end{table}
Table \ref{d R t} shows that the mean radius of the shell is the main geometrical variable, with the mean inter-atomic distance being approximately constant for the various fullerenes C$_n$ of the families $n=60k^2$ and $n=20k^2$ ($k$ integer). The inter-shell separations are approximately constant  for successive fullerenes of each family ($\sim3.5$\AA, \ close to the inter-layer separation in bulk graphite, for the former, and $\sim 2$\AA \ for the latter).
\subsection{Intershell interactions in two-wall fullerenes \label{two-shells}}
\begin{figure*}
\includegraphics[angle=-90,scale=0.5]{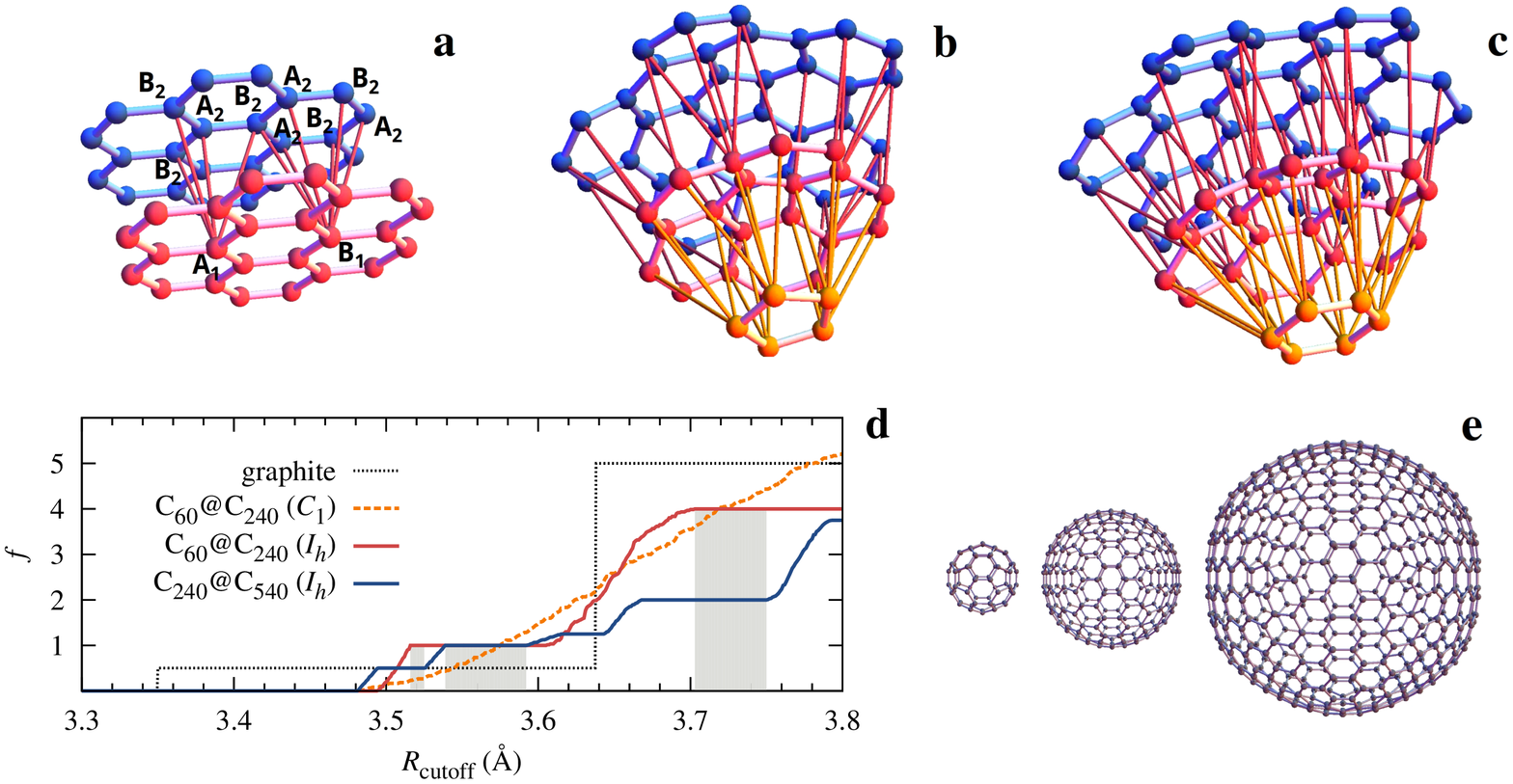}
\caption{\label{bilayer} (a) Scheme of a part of a two adjacent layers of graphite showing the interlayer connections between the lattices A and B from the upper and lower layers, (b)-(c) Schemes of two portions of the three layers of the onions considered in this work and their connections: (b) with the centers of pentagons  aligned, and (c) with the centers of hexagons aligned. These and others configurations occurs in all the multiwall fullerenes considered here, making variable the number of nearest neighbors of the atoms of the shells. (d) Average number of connections $f$ in two-wall buckyonions per atom belonging to the inner sheel as a function of a maximum range for including neighbors $R_{\rm cutoff}$, Eq. (\ref{def f_links}), for C$_{60}$@C$_{240}$ with $I_h$ symmetry,  C$_{240}$@C$_{540}$ with $I_h$ symmetry, C$_{60}$@C$_{240}$ without symmetry (symmetry $C_1$) and for two layers of graphite, for comparison purpose. In graphite only $f=0.5$ and $f=5$ are possible for the range of $R_{\rm cutoff}$ shown, and the step-like variation of $f$ is due to the constrain imposed by the geometrical alignment between the two infinite layers. The three shadowed regions highlight ranges of $R_{\rm cutoff}$ where $f$ has also a step-like behavior for the two-wall onions. (d) Scheme of the single-wall C$_{60}$, C$_{240}$ and C$_{540}$ fullerenes.}
\end{figure*}

We shall study single-wall fullerenes and bilayer and trilayer onions composed by two and three concentric shells of icosahedral symmetry ($I_h$), C$_{n_1}$@C$_{n_2}\ldots$. 
In onions, it has been shown that the resulting structures for each shell are much the same those ones of the isolated fullerenes. Therefore, we assume the geometry for each shell in the onion to be the same as that of isolated C$_n$, taken from Yoshida data base. The most stable mutual orientation is the one that preserves the $I_h$ symmetry of the composite system \cite{Heggie97, Heggie98}. 
Furthermore, in the onion-like family C$_{60k^2}$, it is reasonable to assume the strength of their intershell interactions as similar to those between the layers of graphite, due to the similarity in their interlayer separation.
Figure \ref{bilayer}a shows a scheme of a part of two adjacent parallel layers in graphite separated a distance of 3.35 \AA. Each layer is constituted by two triangular lattices whose sites are denoted as A and B. The subscript 1 or 2 indicates which layer each site belongs. In graphite, the distances and number of nearest neighbors are exactly determined by the relative position between the plane infinite layers of identical geometry. 
In two-wall buckyonions, nevertheless, the inner and outer fullerenes have finite sizes with different number of atoms, different geometrical structures due of their different number of hexagons (e.g., in the family 60$n^2$, every member have 12 pentagons) and eventually different orientations from each other (even though both have $I_h$ symmetry), as shown in Figures \ref{bilayer}b and \ref{bilayer}c. 
Therefore, although it is usually thought of intershell interactions in buckyonions as locally similar to interlayer interactions in bilayer graphene, a more thorough consideration of the intershell hopping is needed due to the differences mentioned above.
%
In graphite the hopping parameter $\gamma_0=t$ describes the covalent bonding arising from $sp^2$ hybridization in a layer.
Inter-layer Van der Waals interactions are described by the parameters  $\gamma_1 = t_\perp\approx 0.4$ eV (hopping energy between atoms A1 and A2), $\gamma_3 \approx 0.3$ eV (between B1 and B2 atoms) and  $\gamma_4 \approx 0.04$ eV (hopping energy for A1-B2 and A2-B1 pairs) \cite{CastroNeto09, Castro10}. 
In bilayer onions, no such two lattices exists due to the finite size and the curvature of the layers. Nevertheless, similar Van de Waals interactions between nearest neighbor (NN) and next-nearest neighbors (NNN) atoms exist. Taking into account that for graphite $\gamma_1\approx \gamma_3$ and $\gamma_4\approx 0$, we take a single value $t_\perp=0.35$ eV for the hopping between pairs of the NN and NNN between layers \cite{Pudlak09}. 
The intershell interaction is included through the hopping integral $t_\perp$ between pairs of NN and NNN atoms as follows: 
we define a cutoff radius $R_{\rm cut off}$ such that every pair of atoms, at ${\bf r}_i$ and ${\bf r}_j$, belonging to adjacent shells and separated a distance shorter than $R_{\rm cut off}$ is assigned a hopping $t_\perp$, i.e.,
\begin{equation}
t_{ij} = \left\{ 
\begin{array}{cc}
t_\perp, & |{\bf r}_i-{\bf r}_j|\le R_{\rm cut off} \\
0      , & {\rm otherwise}.
\end{array}
\right.
\label{def t_ij}
\end{equation}
It should be noted that, due to the faceting of the icosahedral symmetry, the number of intershell connections is not isotropic. 
As an illustration, Figures \ref{bilayer}b and \ref{bilayer}c show two portions of a trilayer onion within a solid angle around the directions joining the centers of the pentagons and hexagons, respectively. The lines joining atoms in adjacent shells represent the intershell connections $t_{ij}=t_\perp$ for a given cutoff radius. Due to the non sphericity of the shells, the outermost pentagon is not connected but the outermost hexagon it is.
We characterize the number of pairs of atoms connected at a given $R_{\rm cutoff}$, by the mean number of neighbors (in the outer shell) `felt' by atoms in the inner shell, 
\begin{equation}
f(R_{\rm cutoff})=N_{\rm connect}/N_{\rm inner},
\label{def f_links}
\end{equation}
where $N_{\rm connect}$ is the number of intershell connections for a given $R_{\rm cutoff}$, and $N_{\rm inner}$ is the number of atoms in the inner shell.
Figure \ref{bilayer}d shows the variation of $f$ with $R_{\rm cut off}$ for C$_{60}$@C$_{240}$ with $I_h$ symmetry,  C$_{240}$@C$_{540}$ with $I_h$ symmetry, C$_{60}$@C$_{240}$ without symmetry (symmetry $C_1$) and for two adjacent layers of graphite. In the latter, there are wide ranges of $R_{\rm cut off}$ where the number of connections keeps constant. Thus, due to the parallel orientation of the infinite layers, there are only some discrete values at which the number of connections increases due to the inclusion of neighbors farther to a given atom. For the onions with $I_h$ symmetry, 
there are common ranges of $R_{\rm cutoff}$ that show step-like behavior for both onions (see regions shadowed in Figure \ref{bilayer}d). In the next section we present results of calculations with a number of intershell connections (characterized by $f$) chosen in those regions. Such a step-like dependence is in contrast to the approximately linear dependence for C$_{60}$@C$_{240}$ with symmetry $C_1$, when the symmetry axes of the $I_h$ fullerenes are misaligned. 
The faceting of the fullerenes induced by the $I_h$ symmetry is particularly noticeable for the largest layers, as can be seen en Figure \ref{bilayer}e for C$_{60}$, C$_{240}$ and C$_{540}$. Therefore, the fraction of carbon atoms having a NN or NNN within the range $|{\bf r}_i-{\bf r}_j|\le R_{\rm cut off}$ is smaller for the larger fullerenes. 
Hence, the TB Hamiltonian for the onions becomes that of the isolated layers with an inter-layer interaction term
\begin{equation}
H_{\rm onion} = \sum_n^{N} H_n + t_\perp \sum_{\langle ij\rangle} (c_i^\dagger c_j + c_j^\dagger c_i ),
\label{H_onion}
\end{equation}
where site $i$ belongs to a inner shell, and site $j$ is its  NN or NNN on the adjacent outer shell.
\subsection{Electronic transport}
When a molecule is attached between two metallic leads and subject to a potential bias, the charge current flowing through it can be calculated with the Landauer equation \cite{Landauer86}
\begin{equation}
I=\frac{2e}{h}\int dE \ T(E) \left[f_L(E)-f_R(E) \right],
\end{equation}
where $f_L$ and $f_R$ are the Fermi distributions at the left (L) and right (R) leads. At low temperatures, the transmission function represents the dimensionless conductance (in units of the quantum $e^2/2h$) and is calculated as
\begin{equation}
T(E) = 4{\rm Tr}({\bf \Gamma}^L {\bf G}^r(E) {\bf \Gamma}^R {\bf G}^a(E)),
\end{equation}
where ${\bf G}^a$ and ${\bf G}^r$ are the matrix representation of the advanced and retarded Green functions ${\bf G}^{r,a}=(E {\bf 1}-{\bf H}\pm i0)^{-1}$, and ${\bf \Gamma}^L$ and ${\bf \Gamma}^R$ are the spectral densities of the leads \cite{Cuevas10}.

In the wide band approximation, the Green function of the connected system can be obtained by using Dyson equation, as
\begin{equation}
G_{1n}^r = \frac{g_{1n}}{1-\Gamma^2(g_{11}g_{nn}-|g_{1n}|^2)-i\Gamma(g_{11}+g_{nn})}.
\label{connected G1n}
\end{equation}
where $g_{ij}$ is the retarded Green function of the isolated system, and $\Sigma_L=\Sigma_R=i\Gamma$ are the self-energies of the leads, considered to be energy independent. Throughout this work, we connect the fullerenes to the leads, using $\Gamma=0.05$ eV, through two carbon atoms located diametrally opposite to each other, one atom in the vertex of a pentagon and its corresponding one obtained by applying the operation of inversion with respect to the center of symmetry.
\section{Results and discussion}
\subsection{Effect of the relative orientation between adjacent shells}
\begin{figure*}
\includegraphics[scale=0.9]{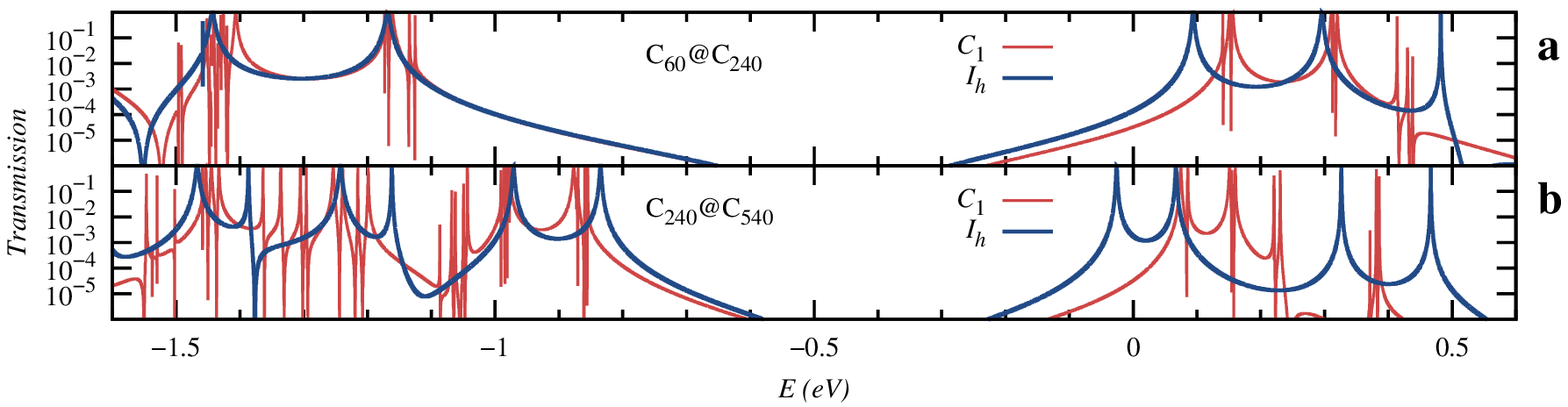} 
\vspace{-0.0cm}
\caption{\label{orientation} Transmission function for (a) C$_{60}$@C$_{240}$, and (b) C$_{240}$@C$_{540}$ buckyonions for the concentric shells rotated an arbitrary angle with respect to each other ($C_1$ symmetry), and with their symmetry axes aligned ($I_h$ symmetry).}
\end{figure*}

The approximately spherical form of the fullerenes, particularly the smaller ones, allows to treat onions as a family of concentric spherical shells for the calculation of some properties, such as the determination of radii of equilibrium, intershell distances, static polarizability or photoionization cross section \cite{Xu96, Ruiz04, Kidun06, Dolmatov08}. For other applications, however, a more accurate description of their geometry is relevant. 
In particular, the trend in larger fullerenes to approach faceted icosahedral forms makes very relevant the relative orientation of adjacent layers, even for concentric shells having individually  $I_h$ symmetry, as already shown in Figure \ref{bilayer}d for the average number of connections per atom $f$.  
The influence of the relative orientation on the quantum transmission is shown in Figure \ref{orientation} for two bilayer onions:  C$_{60}$@C$_{240}$, and C$_{240}$@C$_{540}$. Two orientations were considered: one where the onion has overall $I_h$ symmetry, that is, with the symmetry axes of the individual shell aligned; and one where both shells are concentric but the individual symmetry axes are rotated an arbitrary relative angle from each other ($C_1$ symmetry).
The differences in $T(E)$ between the icosahedral ($I_h$) and the non symmetrical ($C_1$) onion are visible in Figure \ref{orientation}. In the onion with misalignment  between the shells, some degeneracies in the energy spectrum are broken, as reflected in the occurrence of multiple resonant states with many peaks and antiresonances close to each other leading to a trend to the formation of narrow bands for the larger onion C$_{240}$@C$_{540}$. The transmission function of the non symmetrical onions have rapid variations from perfect to vanishing transmission with slight variations of the Fermi energy $E$, while $T(E)$ for the $I_h$ onions is a well behaved smooth function with a few peaks of perfect transmission in the range around the HOMO-LUMO gap. The gap itself increases when the shells become disoriented from each other what also corresponds to a less stable configuration, as previously reported \cite{Heggie97}.
\subsection{Dependence on the number of intershell hopping connections \label{connections}}
\begin{figure*}
\includegraphics[scale=0.90]{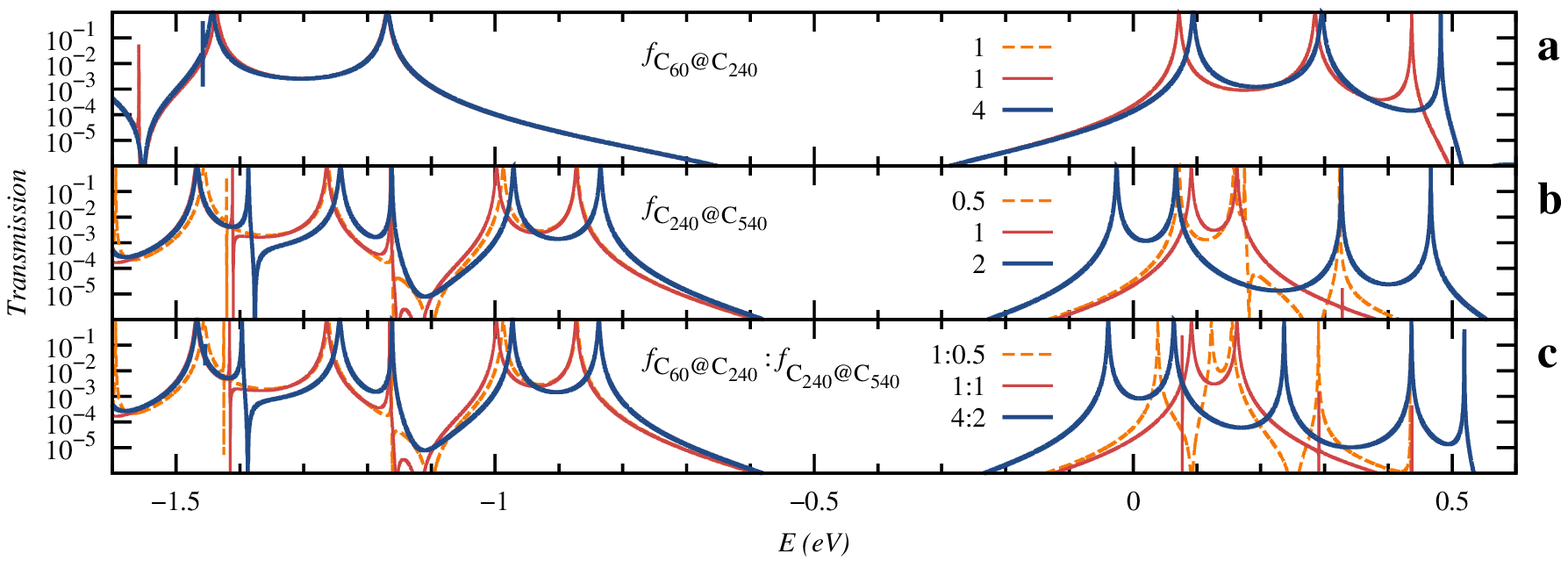} 
\vspace{-0.1cm}
\caption{\label{R_cutoff} Transmission function $T(E)$ of the bilayer and trilayer onions (a) C$_{60}$@C$_{240}$, (b) C$_{240}$@C$_{540}$ and (c) C$_{60}$@C$_{240}$@C$_{540}$ for three different cutoff radii chosen in the shadowed regions of Figure \ref{bilayer}. The resulting mean number of intershell connections $f$, Eq. (\ref{def f_links}), are indicated in each curve.}
\end{figure*}

As mentioned in Section \ref{two-shells}, the relative orientation of hexagons belonging to adjacent shells, the choice of the cutoff radius for defining NN and NNN sites of a given atom and the faceting of the larger fullerenes preclude a unique definition of the number of intershell connections to be included in the Hamiltonian (\ref{H_onion}). Both the interlayer distance in graphite and intershell distance in the $60 k^2$ family are close to 3.5 \AA. Thus, we took two cutoff radii close to this value and a larger one (see shadowed regions of Figure \ref{bilayer}d), to include a bigger number of NN and NNN pairs in the intershell interaction Hamiltonian.
In the following we show the results of calculated $T(E)$ for icosahedral onions ($I_h$@$I_h$) and the aforementioned three choices of $R_{{\rm cutoff}}$. In those three regions, $f=1$, 1 and 4 for C$_{60}$@C$_{240}$, and $f=0.5$, 1 and 2 for C$_{240}$@C$_{540}$. 
In Figure \ref{R_cutoff}, the transmission $T(E)$ is depicted for the three values of $f$, as indicated in the legends. It should be noticed that for the smaller onion C$_{60}$@C$_{240}$, the three $f$ values give almost the same curve (Fig. \ref{R_cutoff}a). 
Therefore, this increase of connectivity does not affect the transmission throughout the composite system. Therefore, the main paths of transmission between shells are along the pairs formed by the atoms of the inner shell with its closest neighbor in the outer one.
Figure \ref{R_cutoff}b shows that the dependence of the transmission on the number of intershell connections is more important for the larger onion C$_{240}$@$C_{540}$. 
For larger fullerenes the deviation from the spherical  shape becomes more noticeable as the number of atoms increases. Thus the C$_{240}$ shell is less spherical than C$_{60}$, and C$_{540}$ is clearly faceted. This departure from sphericity favours the hopping from the inner to the outer shell along certain directions, namely, those in which the icosahedral faces of both shells are closer to each other. Therefore, the larger the number of intershell links the better the quantum transmission. Interestingly, the increasing in the number of connections mainly affect the states above the highest occupied-lowest unoccupied (HOMO-LUMO) gap, noticeably the LUMO and LUMO+1 ones.

Finally, Figure  \ref{R_cutoff}c shows $T(E)$ for the trilayered onion C$_{60}$@C$_{240}$@$C_{540}$ which is notoriously similar to the one of the bilayer onion of Fig \ref{R_cutoff}b, thus showing that the two external shells are the most relevant for the transmission, with only small corrections coming from the innermost C$_{60}$ shell. This effect is discussed in greater detail in Section \ref{shells}. 
The bilayer onion have four peaks above the gap in range shown, namely, those corresponding to LUMO, LUMO+1, LUMO+2 and LUMO+3. The two central peaks (LUMO+1, LUMO+2) are almost degenerates in C$_{240}$@$C_{540}$ but become better resolved in the trilayered onion.
\subsection{Influence of the number of onion shells \label{shells}}
We shall show here that the outermost shell greatly determines the most relevant features of the transmission spectra of bilayered and trilayered onions. The importance of the influence of the inner shells on $T(E)$ decreases inwards.
In Figure \ref{R_cutoff}c we observed that the most noticeable effects when adding C$_{60}$ to the two external shells are the occurrence of an antiresonance between the LUMO and LUMO+1 peaks, and the widening of the narrow Fano-like profile after the LUMO+2 peak (at $E \sim0.2$ eV.). Such a type of effects are analogous to the modifications to the transmission function through a chain introduced by adding a lateral site or chain to the system.
\begin{figure*}
\includegraphics[scale=0.90]{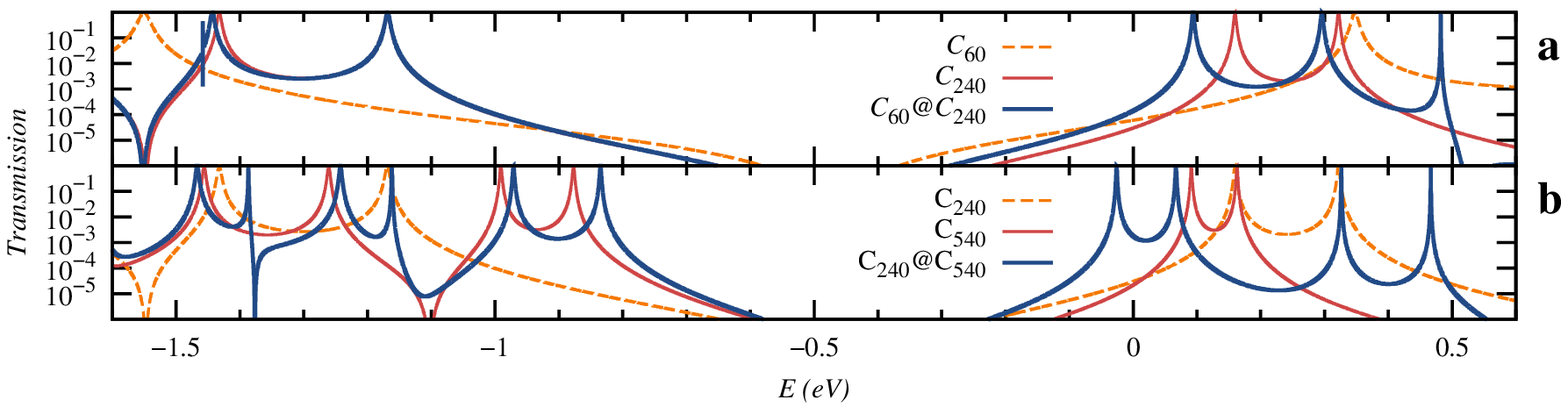} 
\vspace{-0.0cm}
\caption{\label{shell number} Transmission function for (a)C$_{60}$@C$_{240}$ and (b) C$_{240}$@C$_{540}$ compared to those for the single-wall fullerenes.}
\end{figure*}
In the following we fix the number of connections in trilayer onions by chosing $f=4$ ($f=2$) between the two outermost (innermost) shells.
Figure \ref{shell number} shows the transmission for the bilayer onions C$_{60}$@C$_{240}$ and C$_{240}$@C$_{540}$ as compared to those for the corresponding single-wall fullerenes. It can be seen that for energies below the gap, $T(E)$ for the onions is very similar to the one of the outer fullerene, i.e., the one for C$_{240}$ in Figure \ref{shell number}(a), and for C$_{540}$ in Figure \ref{shell number}(b). 
The peaks above the gap preserves similar features both in  C$_{60}$@C$_{240}$ and C$_{240}$, although with a relative energy shift of the peaks.
\begin{figure*}
\includegraphics[scale=0.90]{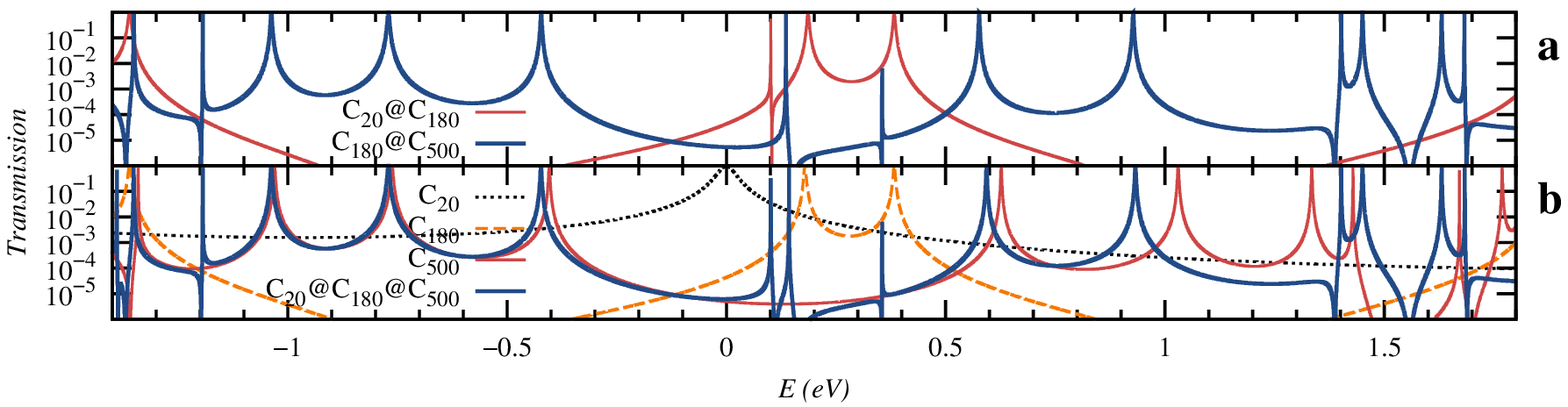} 
\vspace{-0.0cm}
\caption{\label{family 20} transmission function for (a)C$_{20}$@C$_{180}$ and C$_{180}$@C$_{500}$, and (b) C$_{20}$@C$_{180}$@C$_{500}$ compared to those for the single-wall fullerenes.}
\end{figure*}
In Figure \ref{shell number}b similar considerations can be made about the comparison the transmission through C$_{240}$@C$_{540}$ and C$_{540}$; the latter provides most of the features observed in the former, particularly for energies below the gap. The intershell connections in C$_{240}$@C$_{540}$ eventually contributes to the occurrence of antiresonances, such as that at $\sim-1.4$ eV, not present in $T(E)$ for C$_{540}$.
In other cases, it softens the vanishing of transmission, such as in the antiresonance of C$_{540}$ at $\sim-1.1$ eV which becomes in a finite transmission for C$_{240}$@C$_{540}$. 
Roughly speaking, introducing C$_{60}$ as a third innermost shell does not greatly modify the transmission of the bilayered onion.
The influence of the external shell on the transmission can be interestingly shown in the onions formed from fullerenes belonging to the family $20 k^2$, which also have icosahedral symmetry. Thus, C$_n$ fullerenes with $n=20$, 80, 180, 320, $500\ldots$ are predicted to be stable, with a radii increasing by approximately 2 \AA \ from each member from the family to the next one. The radii of equilibrium for the family $60 k^2$ were shown to be accurately determined by a continuous spherically symmetric Lennard-Jones model \cite{Baowan07}. Application of the same model for the multiwall onions of the $20 k^2$ family (with $k$ odd integer), results an intershell distance of about 4 \AA, not far from the intershell distance for the $60 k^2$ family or the interlayer distance in graphite. Therefore, we show calculations for the single-wall, two-wall and three-wall fullerenes obtained from C$_{20}$, C$_{180}$ and C$_{500}$.
Our TB
calculations show that C$_{20}$ and C$_{500}$ are gapless, while C$_{180}$ presents a gap of $\approx 1.5$ eV. In Figures \ref{family 20}a and \ref {family 20}b, C$_{180}$ is the outer and inner shell respectively. As a consequence, in Figure \ref{family 20}a, C$_{20}$@C$_{180}$ shows a region of vanishing transmission within the gap, while it keeps finite for C$_{180}$@C$_{500}$, except for the already discussed narrow antiresonances arising from the intershell connections. Hence, the transmission function is strongly sensitive to the electronic structure of the external shell. As a final example of this property, Figure \ref{family 20}b, shows the transmission for the trilayered C$_{20}$@C$_{180}$@ C$_{500}$ as compared to those of the single-wall fullerenes. It is seen that the most of the features of the onion are reproduced by the transmission through C$_{500}$, with some Fano-like antiresonances originated in  the transmission spectrum of the inner C$_{180}$ and, to less extent, in the innermost C$_{20}$.
It can bee seen that the three-wall onion (Figure Figure \ref{family 20}b) is well described by the two-wall one (Figure \ref{family 20}a). Interestingly, the Fano-like resonance of C$_{20}$@C$_{180}$@ C$_{500}$ at $E\approx 0.1$ eV is not present in C$_{180}$@ C$_{500}$ but it is in C$_{20}$@C$_{180}$; therefore, such a peak of conductance is an effect from the coupling of the two inner sheells.
\section{Conclusions}
In this work we have studied theoretically the quantum transmission through single-wall fullerenes and bilayered and trilayered onions of icosahedral symmetry, when attached to metallic leads, by using a TB
Hamiltonian and Green functions methods. Although the Van der Waals interactions between onion shells are supposed to be similar to those between graphite layers, the finite size of the fullerenes, their curvature and relative orientations need some analysis for including the intershell hopping parameter. We include in the model the effect of finite size and curvature through a parametrization of the hopping integral as a function of the number of atoms of the shell. The number of connections from a given atom to others belonging to adjacent shells was studied by introducing a cutoff radius for the interaction. 
We found that misalignment of the symmetry axes produces breaking of the level degeneracies of the individual shells, giving rise some narrow quasi-continuum bands instead of the localized discrete peaks of the individual fullerenes. Most of the features of the transmission through the onions are already visible in the transmission function of the single-wall fullerene forming the outer shell. The main modifications between them are antiresonances arising from the coupling between the outer layer with the next innermost one. For three-wall onions, the transmission becomes barely sensitive to the most internal shell. Interestingly, when the fullerene of the external shell is gapless, the transmission of the onion does not vanish along finite ranges of energy. This property could be useful for designing multilayered fullerenes with tailored conductance by properly growing the outermost layers.
\section*{Acknowledgments}
We acknowledge financial support for this project from CONICET (PIP 11220090100654/2010) and SGCyT(UNNE) through grant PI F007/11.

\end{document}